\documentclass[preprint,12pt]{aastex} 
\usepackage{graphicx}
\newcommand{\mapiii}{\textsc{Mappings iii}}
\newcommand{\kms}{\ensuremath{\,\mbox{km}\,\mbox{s}^{-1}}}

\newcommand{\heii}{He\,{\sc ii}}

\newcommand{\ciii}{C\,{\sc iii}}
\newcommand{\civ}{C\,{\sc iv}}
\begin{document}
\title{Spatially Resolved Ultraviolet Spectra of the
High-Velocity Nuclear Outflow of NGC 
1068\footnote{Based on observations made with the NASA
\emph{Hubble Space Telescope}, obtained at the Space Telescope Science
Institute, which is operated by the Association of Universities for Research
in Astronomy, Inc., under NASA contract NAS5-26555. These observations
are associated with proposal ID GO-7353.}} 
\author{Brent A.\ Groves} \affil{Research School of Astronomy \& Astrophysics,
Australian National University, Cotter Road, Weston Creek, ACT 2611 Australia}
\email{bgroves@mso.anu.edu.au}
\author{Gerald Cecil} \affil{Dept.\  of Physics \& Astronomy, University of
North  Carolina at  Chapel  Hill, Chapel  Hill,  NC, USA  27599-3255}
%
\author{Pierre Ferruit} \affil{Observatoire de Lyon, 9 Avenue Charles Andre,
Saint-Genis, Laval Cedex F-69561, France}
\author{Michael  A.  Dopita}  \affil{Research  School of  Astronomy  \&
Astrophysics, Australian National University, Cotter Road, Weston Creek, ACT
2611 Australia}
\begin{abstract}
We present ultraviolet emission-line maps of the narrow-line region
(NLR) of NGC 1068. The maps span 115--318 nm, the biconical ionization
cone, several posited jet/ISM interactions, and the compact knots
whose optical spectra we reported previously resemble kinematically
the quasar Associated Absorption Line systems.  Across the NLR, we
find that ultraviolet flux ratios are consistent with photoionization,
not shock excitation, even for gas blueshifted abruptly to $3000$
\kms\ relative to galaxy systemic velocity or for gas projected near
the radio jet.  The knots may be radiatively accelerated, photoablated
fragments of molecular clouds.
\end{abstract}
\keywords{galaxies: active --- galaxies: individual (NGC 1068) --- galaxies:
kinematics \& dynamics --- galaxies: Seyfert}
\section{Introduction}

While broad emission line profiles are a characteristic signature of
the energetic processes in the center of active galaxies, the spectra
of some systems also show absorption lines. Such absorption is
generally found against the ultraviolet (UV) continuum, where the line
profiles can extend blueward of galaxy systemic velocity from a few
hundred to several thousand \kms\ \citep{HamBarJB97}. Whereas the
properties of the emitting clouds are reasonably well understood, less
is known about the absorbing gas, especially how it is accelerated to
such high velocities. The main problem in understanding such systems
is that the background continuum necessary for absorption in active
galaxies is often present only at the nucleus and at distributed,
compact ``hot spots''.  

Space telescopes permit the study of active
galaxies with much better spatial resolution and at UV wavelengths.
Such a study of selected nearby systems can clarify both the
background against which the absorbers are seen, and --- the subject
of this paper --- the connection between the emitting and the
absorbing gas.  In \citet[hereafter Paper I]{Cecil02} we reported on
results derived from a grid of Space Telescope Imaging Spectrograph
(STIS) medium resolution spectra that cover much of the spatially
extended narrow line region (NLR) emission of the nearby Seyfert
galaxy NGC 1068. As Figure \ref{oiii1068} shows, we found that many
compact ``knots'' in this NLR span velocities greater than 3000 \kms\
in radial velocity, and are consistently blueshifted from galaxy
systemic.  The profiles of their optical spectral lines resemble those
of associated absorption line (AAL) systems seen in some quasar
spectra. Using combined ground--based optical/infra-red and UV
emission-line spectra, we can constrain the internal physical
properties of these objects. In this paper we discuss the UV spectra.

UV spectra are potentially more sensitive probes of NLR conditions
than lines in the visible, because collisionally excited UV lines are
strong when emitted from the cooling region behind high-velocity
($150-500$ \kms) shocks.  \citet[ADT hereafter]{ADT98} showed that the
ratios of UV line fluxes can even discriminate between the two major
excitation mechanisms posited for NLRs: shock fronts moving at
$\la400$ \kms\ with photoionized precursors, and gas that has been
photoionized by the AGN non-stellar continuum.  In practice, UV
spectra have been of limited use in nearby objects because mapping
extended NLRs even at low spectral resolution is arduous with current
space spectrometers, and the emission is easily extinguished by dust.
While NLRs are widely assumed to be photoionized structures, patterns
of their internal motions suggest three means of gas acceleration:
virial in the gravitational potential of the galaxy bulge
\citep{ne96}, locally at the boundaries of expanding radio lobes
\citep[for example]{ax98}, and radial outflow from the nucleus
\citep[for example]{CK00}.  In the case of NGC 1068, previous,
sparsely sampled long-slit spectra \citep{ax98,KC00a} found
discontinuities in the velocity field at several points that seem to
coincide with excitation changes, suggesting to those authors a local
role for shocks.  \citet{gr99} argue that shocks are responsible for
the bright \ion{C}{3}$\lambda$977 and \ion{N}{3}$\lambda$991 emission
in the off-nuclear \emph{HUT} spectra.  On the other hand, spatially
resolved \emph{XMM/Newton} spectra \citep{Kink02} indicate that most
of the X-ray emitting gas throughout the extended NLR is photoionized,
not shocked excited. In addition, photoionization models of a single
spatially-resolved UV/visible long-slit spectrum of this NLR can
accurately reproduce the emission line ratios \citep{KC00a,KC00b}. 

In
this paper, we discuss UV STIS low-resolution spectra that span
$\sim$30\% of this NLR at high (0\farcs05 = 3.5 pc) spatial
resolution.  In \S2 we discuss the acquisition and reduction of the
spectra, and register them to the medium resolution optical spectra
discussed in Paper I to better constrain the physical processes at
work in the NLR.  In \S3 we plot the emission-line flux ratios in
diagnostic diagrams to compare them to the predictions of various
excitation models. In \S4 we discuss the results in terms of the
physical processes occurring in the high-velocity knots, and how these
relate to Associated Absorption Line (AAL) systems in some quasar
spectra.  Our conclusions are summarized in \S5. We assume that the
distance and systemic velocity of NGC 1068 is 14.4 Mpc and $cz =
1148~\kms$.

\section{Observations \& Reductions}

Seven UV spectra of NGC 1068 were recorded on the \emph{HST} STIS/MAMA
detectors: a pair of FUV+NUV exposures at three parallel slits placed
along the axis of the radio jet, and a single NUV exposure at a fourth
parallel position.  Our observations were detailed in Table 1 of Paper
I.  

Figure \ref{oiii1068} registers the STIS slits to the [\ion{O}{3}]
clouds and radio image \citep{ga96}; UV spectra were recorded at
[\ion{O}{3}] slit positions 2--4 (and 5 for the fourth NUV slit).
These positions primarily sample the high-velocity, blueshifted knots
discussed in Paper I, and secondarily sample gas on and adjacent to
the radio jet. Time constraints prevented us from mapping parts of the
NLR farther to the NW where there is optical or radio evidence for
jet/ISM interactions, in particular at clouds C (covered by slit 5,
and hence only has an NUV spectrum) \citep[see][]{ga96}, G
\citep[see][]{pe97} (also covered by slit 5), and H
\citep[see][]{ax98} (covered by slit 6, and hence has no UV
observations).  

The 0\farcs2-wide slit (0\farcs19 on the sky) that we
used sacrifices velocity resolution for spatial coverage.  This slit
projects to a width of 8 pixels on the MAMA detectors and produces
\footnote{See 
www.stsci.edu/hst/stis/documents/handbook/currentIHB/images/c13\_specrefa30.gif
} 
``triangular'' wings on narrow emission lines that span 10 pixels.
The UV PSF of the telescope within the STIS slit is much smaller than
this width. Hence, compact emission knots at the same radial velocity
will map to different wavelengths in the dispersed spectrum.
STIS/MAMA dispersion is 0.6 \AA\ and 1.58 \AA\ per pixel in the G140L
and G230L gratings used for FUV and NUV spectra, respectively. This
implies maximum velocity shifts from nominal positions for two knots
separated by the slit width of $\pm465$ (ie. 930 \kms\ total width) and $\pm375$ \kms\ respectively (see www.stsci.edu/hst/stis/design/gratings).
Lacking narrow-band images of the NLR in all spectral lines, we cannot
map knots to specific locations across the slit. For our purposes,
this significant ``slit effect'' uncertainty is nonetheless tolerable
because the components evident in the emission-line profiles in Figure
\ref{profiles} range over much larger velocities.  Certainly, without
the medium resolution spectra to guide us, it would have been
impossible to extract reliable excitation trends from the low
resolution spectra.

\placefigure{oiii1068} 

\placefigure{UVspec} 

The STIS data processing pipeline delivered wavelength and flux calibrated
spectra as well as the statistical error in flux at each pixel. We
removed continuum light by averaging across line-free intervals,
fitting the result with a low-order Chebyshev polynomial, and
subtracting the fit from the data.  To deredden spectra, we used the
standard reddening curve of \citet{MRN77} and estimates that KC
derived from the \ion{He}{2}$\lambda 1640/\lambda 4686$ flux ratio
along their single slit.  KC found systematic differences between
blue-- and red--shifted emission, with gas in the NE--blue quadrant
more heavily reddened ($E_{B-V}\sim 0.35$) than that in the NE--red
quadrant ($\sim0.22$).  Although they placed their slit along
P.A.~22\arcdeg, rather than our 38\arcdeg, and displaced it 0\farcs14
north of the continuum peak, we had to apply their values to all of
our spectra because we could not include the \ion{He}{2}$\lambda 4686$
line in our spectral maps.  Figure \ref{UVspec} shows one of the three
resulting FUV+NUV spectrograms.  The blue wing of the strong Ly
$\alpha$ line is invariably obliterated by geocoronal emission, so is
ignored in this Figure and in our analysis.  

To improve signal to
noise ratios, we binned spectra into the spatial regions marked in
Figure \ref{oiii1068} and into velocity increments of 200 \kms. These
intervals were chosen to bracket distinct changes in kinematical
behaviour and to encompass discrete radio knots.  To link to our
previous analysis, we rebinned the [\ion{O}{3}] and H$\beta$ spectra
from Paper I to the spectral/spatial resolutions of the UV spectra;
the two line sets are registered in Figure \ref{UVspec}.  Figure
\ref{profiles} compares the line profiles extracted from each
spectrogram at each of these regions.  Regions 1 and 2 contain the
brightest emission from the base of the SW ionization cone, region 3
encompasses the nucleus and the base of the NE ionization cone so
shows the brightest lines, while regions beyond show the profiles from
kinematically distinct regions across the cone.

\section{Results}

Our spectra show the UV lines usually visible from a moderately
reddened NLR, especially strong [\ion{Ne}{4}] and \ion{C}{4} nearest
the nucleus.  Note that the high-velocity features on the detailed
optical spectra in Figure \ref{oiii1068} are also prominent on the UV
lines --- especially on the strongest three, \ion{He}{2}$\lambda
1640$, \ion{C}{4}$\lambda \lambda 1549$, and \ion{C}{3}]$\lambda
1909$.  Each panel in Figure \ref{profiles} compares these profiles at
one spatial region and slit position.  Slit 4, panel 3 covers the
``continuum hotspot'' discussed by \citet{cr00}, and confirms the
broad \ion{C}{4} line profile there, which \citet{an94} found is
polarized because it reflects emission from the hidden BLR.  As
discussed in Paper I and in \citet{cr00}, the spectra of many bright
knots in other regions of the NLR also show evidence for a few percent
contribution from nuclear light.

The axis of the ionization bicone lies close to the plane of the sky
\citep[for example]{cbt}, about 45\arcdeg\ above the galaxy disk.  As
well as picking up scattered nuclear light, our line of sight
therefore penetrates a large range of gas densities and, plausibly,
ionization conditions through the galaxy atmosphere, depending on
whether we see gas in front of the jet (i.e.\ accelerating above the
galaxy disk) or below it (i.e.\ decelerating into the denser galaxy
disk).  Line profiles throughout the NE cone have consistently
stronger blue wings, suggests that NLR clouds are being pushed
laterally from the cone axis, toward us.

\placefigure{profiles}

\subsection{Models of Gaseous Excitation}

We summarize the spatio-kinematic variations in excitation in the
diagnostic diagrams of ADT, plotting the ratios of \ion{C}{4} $\lambda
\lambda 1549$/\ion{He}{2} $\lambda 1640$ against \ion{C}{4} $\lambda
\lambda 1549$/\ion{C}{3}] $\lambda 1909$ (figure \ref{UVratios}).
These have been chosen because the three are among the strongest
lines, their ratios are least affected by reddening uncertainties and
the contribution of scattered nuclear light, and, according to ADT,
most clearly distinguish between photoionization and shock excitation.
We used Monte Carlo techniques to estimate errors from the combined
uncertainties of the lines fluxes, the continuum subtraction, velocity
rebinning, dereddening (assuming a bump in its curve near $\lambda$218
nm), and the division of the different line profiles to form ratios.
While there is evidence that reddening in this NLR is anomalous
compared to both the Milky Way and the Magellanic Clouds \citep{an94},
and dust is presumably also distributed within the densest clouds, the
ratios of ADT are chosen to be close in wavelength to minimize such
uncertainties.  The arrow on panel (1) of figure \ref{UVratios} shows
the effect of an additional external extinction of $3~A_V$ upon the
models, which is however unwarranted by existing observations.

\placefigure{UVratios}

Figure \ref{UVratios} plots the observed flux ratios at various points
in the NLR with the predictions of both shock and photoionization
models.  Data colors and symbols are explained in the plot caption.
Each panel a--c is bisected, with the left panels comparing our
spectra with shock models, and the right with photoionization models.

\subsubsection{Shock Models}
We compared the data to two grids of shock models from ADT: a pure shock
only, and the shock plus its photoionizing precursor; \citet{DS96}
detail the model physics.  Grids are labeled by shock speed and
magnetic parameter (an ambient magnetic field inhibits shock
compression).  Essentially all of the data points are clearly displaced
from the shock grids.  Increasing extinction by $\sim3~A_V$ beyond
what we have used would bring many points into agreement with the
shock predictions, but there is no evidence for an external dust
screen this thick.

\subsubsection{Photoionization Models}
We used the latest version of the shock/photoionization code \mapiii\
\citep{Groves04a}.  Density and the spectral index $\alpha$ of the
single ionizing power-law label our photoionization grids; using a
structured, filtered ionizing continuum such as that of \citet{al00}
produced similar results for this application.  The final input is the
ionization parameter $U$, which measures the number of ionizing
photons per hydrogen atom at the inner surface of the cloud.  The
right-hand panels of Figure \ref{UVratios} show three sequences of
photoionization models:
\begin{enumerate}
\item A sequence of constant density, isochoric, ionization bounded clouds.
\item The A$_\mathrm{M/I}$ sequence of \citet{Binette96}, which
follows by combining the emission from
matter bounded and ionization bounded clouds. The
sequence is parameterized by the ratio of the solid angle occupied by the
Matter bounded component to that of the Ionization bounded component.
\item A sequence of dusty, radiation pressure dominated clouds
from \citet{Groves04a,Groves04b}.  Each cloud is pressurized mostly by
radiation while being exposed to the same input spectrum as the
isochoric model.  The cloud is therefore isobaric, so density here
refers to the depth where \ion{H}{2} = \ion{H}{1}, which is approximately where
[\ion{S}{2}] is emitted, and $U$ where the temperature reaches $2
\times 10^4$ K predominantly by photoelectric heating.
\end{enumerate}

\subsection{Line Ratio Diagrams}
Panel a) in figure \ref{UVratios} shows ratios along slit \#2.  The
high velocity ratios (drawn as black ellipses) cluster within the same
region of the line diagnostic diagram. Their \civ/\ciii] ratios are
too low to be reproduced by the shock models, but are in reasonable
agreement with the photoionization models. The only discrepancy is
from region 1, which may be due to low \civ\ flux and relatively high
noise. The systemic ratios (drawn as green ellipses) are more
dispersed, with some in the same region as the high-velocity clouds
while others closest to the nucleus (2 and 3) have high \civ/\ciii]
ratios. The regions with high \civ/\ciii] ratios are equally well
represented as either shock excited clouds or regions with high
ionization parameter.

Panel b) shows ratios along slit \#3. They are similar to slit \#2, but
some ratios (heavy ellipses) are from points coincident with the radio
jet, as seen in figure \ref{oiii1068}. Compared to panel a), the
high-velocity ratios cluster less tightly and at smaller values of \civ/\ciii]
and \civ/\heii, whereas those coincident with the jet have larger
\civ/\ciii] values. However, most of the high-velocity ratios are
still better reproduced by the photoionization models, with the jet
coincident ratios possibly containing some shock contribution. The
systemic spectra are again dispersed, with the nucleus dominated
spectrum region 2 showing the strongest \civ/\ciii] and \civ/\heii
ratios. The systemic observations with high \civ/\ciii] can possibly
be reproduced by shocks, but most require that photoionization
dominate.

Ratios along slit \#4 are shown in panel c). Once again the
high-velocity emission is clustered, now slightly more dispersed and
with smaller \civ/\ciii] than in panel a). Jet coincident spectra
again show larger \civ/\ciii] ratios.  Although the photoionization
models are closer than the shock models to the observations, the data
are still significantly displaced from the models. This could be due
to much greater internal reddening, or different cloud parameters such
as greater metallicity.  The systemic ratios in slit \#4 are more
clustered than those in previous slits. Regions 2, 3, and 4, which
coincide or are close to the nucleus, have smaller values of
\civ/\ciii] than the other systemic observations, differing from the
previous slits.

\section{Discussion}

In each panel of Figure \ref{profiles}, \ion{C}{4} is the strongest
line, indicating a high state of gaseous excitation. In addition, the
emission in each region is brightest around systemic velocity.  For
gas accelerated from rest and excited by a shock front, one would
expect the opposite trend: gas at the \emph{highest} velocities would
be moving closest to the shock, so would be the most highly excited.

In general, data in Figure \ref{UVratios} plot closer to the
photoionization than the shock models.  In particular, the last 3
panels of all slits in Figure \ref{profiles} span the extended NLR
beyond 2\arcsec\ radius, and show similar, red asymmetric line
profiles and photoionized flux ratios.  In contrast, profiles from
region 5 are multipeaked, and this region has enhanced X-ray emission
\citep{yo01}.  The [\ion{O}{3}] image shows that this region is
delineated by clusters of unresolved knots whose [\ion{O}{3}] spectra
(Figure 1) are very blueshifted and have large velocity dispersions.
Despite their kinematical discontinuities, these spectra also have
much smaller \ion{C}{4}/\ion{C}{3}] ratios than produced by the shock
and shock+precursor models in Figure \ref{UVratios}.  This finding is
consistent with the strong signature of photoionization in
\textit{Chandra} spectra \citep{Kink02}, and strong coronal-line
emission evident in optical \citep{KC00a} and IR \citep{Ma96} spectra
of the NE quadrant of the NLR between 1--2\arcsec\ radii.

\subsection{Effect of the Radio Jet}
Comparing Figures \ref{oiii1068} and \ref{profiles}, we sought
correlated changes in gaseous excitation and kinematics that might
signify shocks arising through jet-cloud interactions or compression
along expanding radio lobes.  We spanned the SW half of the jet with
slit 4 (slit 5 spanned its NE half but only in the NUV, preventing us
from deriving interesting line ratios).  Regions 3--5 include some
posited jet/ISM interaction sites.  Across region 3 the [\ion{O}{3}]
profiles in Figure 1 panel (4) straddle systemic velocity.  Although
the [\ion{O}{3}] profiles span in excess of $\pm2000$ \kms, both
carbon ions have a broad blue wing centered around --3500 \kms\ from
systemic velocity; this feature is associated with cloud B and
coincides both with the UV continuum ``hotspot'' of scattered nuclear
light and with the posited jet/ISM interaction of \citet{ga96}.
Despite these large velocities, the UV profiles show excitation
changes only near systemic velocity, where \ion{C}{4} is about
1.5$\times$ stronger than \ion{C}{3}].  In fact, the flux ratios of
the blue and red wings plot among the farthest from the shock grids in
Figure \ref{UVratios}.

\subsection{Radiative Acceleration}
Our more extensive spectra therefore confirm that UV/optical gas in this 
ionization cone is not dominated by shocks.  How, then, are such compact knots
accelerated to high velocities without disintegrating?  In Paper I
\S 4.3.4 we explored radiative acceleration of \emph{dusty} clouds, and
detailed this idea in \citet{Groves02}.  The radiative force
photoablates gas+dust from the surface of the NLR cloud, driving it
away from the ionizing source \citep[see Figure 1
in][]{Groves02}. Because much of the ionizing opacity at high $U$ is
from dust that is coupled tightly to the gas by Coulomb interactions,
this force can be appreciable and really applies to the whole cloud.
At the edges of NLR clouds there is no counteracting pressure
gradient, and the full radiative force accelerates this gas from the
active nucleus.  Such a model is supported by \citet{Ga03}, who
conclude that the dust in the NLR is probably distributed in optically
thick dust clouds.

As noted in Paper I, during its outflow the high-velocity
tail of the NLR clouds will be shielded from ionization by the slower
flow that feeds the ionizing central source.  Figure \ref{UVratios}
shows this ionization gradient: both blue- and redshifted components 
have smaller \ion{C}{4}/\ion{C}{3}] ratios
than the systemic components. This trend would arise if the high
velocity flows see a smaller $U$, perhaps from a filtered continuum
\citep[as posited by][]{al00}, compared to that of the systemic
component.

\subsection{Are the Knots Related to Associated Absorbers?}
We noted in Paper I that, if the high-velocity emission knots were
seen in absorption against the nucleus instead of in emission against
the galaxy disk, then their kinematics would resemble those of
associated absorption line (AAL) systems in quasar spectra. We have
now shown that they have ionization structure similar to those of
AAL's: both have strong UV resonance lines \ion{O}{6}, \ion{N}{5},
\ion{C}{4}, \ion{Mg}{2}, and \ion{Si}{2} \citep[for example for
AAL's]{HamBarJB97}.  The high-velocity knots in this NLR may,
therefore, be resolved AALs seen in emission.

This hypothesis can be tested by determining the column density and
total masses of the knots.  Where measurable and unsaturated, the
columns of AAL's are $N_H=10^{19}$ to $>10^{20}$ cm$^{-2}$ in
moderately ionized regions.  This gas is thought to arise from an
accretion disk wind, and the observed scaling of \ion{C}{4} equivalent
width with increasing UV luminosity (``Baldwin effect'') has been
interpreted as evidence for a radiation--pressure driven outflow
\citep{Veste03}.  Well studied AAL's often show even more highly
ionized X-ray absorption \citep[for example]{ge98,cr99}, but their
distance to the galaxy nucleus is poorly constrained by UV line
variability and excited-state density diagnostics, so results are
highly uncertain.

Present constraints on masses \citep{Cecil02} for the compact knots in
NGC 1068 assume case-B recombination conditions, hence limit only the
ionized mass to typical values 5--10 M$_\odot$/($10^4$ cm$^{-3}$).
However, the blueshifted clouds are 0\farcs5--1\farcs7 NE of the
nucleus, so integral-field optical/IR spectral maps can constrain
cloud columns and total masses in a straightforward and reliable
fashion.  Such spectra would further constrain the physical properties
of the high-velocity NLR clouds by mapping ionized gas density,
temperature, and $U$ gradients, and may potentially clarify properties
of unresolved quasar AAL's and radiatively accelerated outflows in
general.

\section{Conclusions}

Like the [\ion{O}{3}] spectra presented in Paper I, \emph{HST} UV
spectral maps of the NLR of NGC 1068 show bright regions of emission
that have been accelerated abruptly to velocities greater than 3000
\kms\ relative to adjacent gas. Knot excitation is not influenced
directly by the radio jet.  Our photoionization models better
reproduce the flux ratios of the strongest UV emission lines across
the NLR than do shock models.  While these UV spectra do not by
themselves strongly constrain a particular photoionization model,
incorporating other diagnostic flux ratios that are readily obtained
from optical emission-line spectra may do so \citep{Groves04b}.  With
no indication of shock interaction, the high-velocity knots are
certainly predominantly photoionized, and may be radiatively
accelerated, photoablata from NLR clouds.  They show the same
ionization and kinematical structure as the associated absorbers seen
in some quasar spectra, hence may provide us with a unique opportunity
to resolve a fundamental aspect of quasar dynamics: radiative
acceleration of dusty clouds by the AGN.

\acknowledgements 
B.G.~was supported in this collaboration by an Alex
Rodgers Traveling Scholarship. M.D.~acknowledges the support of the
Australian National University and of the Australian Research Council
through his ARC Australian Federation Fellowship. We thank Mark Allen
for providing us with his most recent UV models, and Ralph Sutherland
for useful discussions.

%
%
\begin{figure}[!htp]
\centering
\caption{HST/FOC [\ion{O}{3}] image (\emph{a}) of the NLR of NGC 1068
\citep{ma94}.  The vertical scale at right is in arcseconds from the
nucleus (1\arcsec\ = 70 pc) and runs along P.A.\ 38\arcdeg\ with NE at
top. Slits used for optical (Paper 1) and FUV+NUV spectra 
(e.g.\ Figure \ref{UVspec}) are
shown and numbered at top.  The regions from which spectra in
Figs.\ \ref{profiles} \& \ref{UVratios} are extracted are numbered on the left, with the white
horizontal lines marking the division between regions in each slit. The radio jet is indicated by the contours. The letters and ellipses mark compact knots and high-velocity [\ion{O}{3}] features respectively and are discussed in greater detail in Paper 1.
The 3 panels (\emph{b})
show the [\ion{O}{3}] emission-line profiles for slits 2--4 shown on
(\emph{a}) that have been
discussed in detail in Paper I. The markings on the right show the vertical scale in arcseconds (as in \emph{a}) and the regions from which the UV spectra are extracted. Also marked are the regions and radio jet from (\emph{a}) which are contained within the relevant slit. The horizontal scale indicates the velocity of the various features from NGC 1068 systemic velocity (v$_\mathrm{sys}$).\label{oiii1068}}
\end{figure} 
\begin{figure*}[!htp]
\end{figure*}

\begin{figure}[!htp]
\caption{A continuum subtracted, STIS FUV+NUV spectrogram of NGC 1068, 
this one adjacent to the radio jet (panel 3 in Figure \ref{oiii1068}), 
using the 0\farcs2-wide slit.  The horizontal lines are those of Figure
\ref{oiii1068}, and delineate the regions over which we averaged spectra
(and are shown before continuum subtraction).
The H$\beta$ and [\ion{O}{3}] profiles from Paper I 
have been rebinned to match the resolution of the UV
data, and are shown before continuum subtraction.
High-velocity features are evident on many of the lines. To fit
everything on this plot, the wavelength scale is linear but discontinuous,
and the intensities are log-scaled.\label{UVspec}}
\end{figure}

\begin{figure}
\caption{Emission-line profiles of (green) \ion{He}{2}, (black) \ion{C}{4},
and (red) \ion{C}{3}, are shown dereddened but otherwise unscaled. In
the grid of spectra, slit number is vertical, and spatial interval
across the NLR is horizontal and increases from SW at left region 1,
the AGN in region 3 (of slit 3), into the NE cone toward the right.
Small ticks are every 1000 \kms\ in the full range of $\pm$9000 \kms,
centered on galaxy systemic velocity.\label{profiles}}
\end{figure}

\begin{figure*}
\end{figure*}

\begin{figure}
\caption{Flux ratios from dereddened STIS spectra across the NLR
are compared to the results of (left columns) shock and (right
columns) photoionization models. Panel a) contains the regions from
slit (2), b) slit (3) and c) slit (4). The $2\sigma$ error ellipse
from each pair of line ratios is numbered by its extraction region in
Figure 1.  Heavy ellipses correspond to the ratios at points
coincident with the radio jet, while light ellipses corresponds to
points off the jet.  The
black ellipses plot ratios of high-velocity gas
(beyond $\pm500$\kms\ of galaxy systemic), whereas the green plot
ratios
for velocities within $\pm500$\kms\ of systemic.  The two models from
ADT shown in the left-hand panels are for a pure shock and a
shock+photoionizing precursor; both are labeled by the shock speed and
magnetic parameter $B/\sqrt{n}$.  Three photoionization models are
shown in the right panels: the $A_{M/I}$ sequence varies the covering
fraction of matter- to ionization-bound clouds. The other curves plot
at top a single isochoric cloud, and at bottom an isobaric, dusty
radiation pressure dominated cloud. Each model varies ionization
parameter $U$ from $\log U=-3.0$ to 0 in steps of 0.3 from right to
left, at two values of the gas ([\ion{S}{2}]) density $n=10^2$ \&
$10^4$ cm$^{-3}$ in the case of the dusty model.
\label{UVratios}}
\end{figure}

\end{document}